\documentclass[simplecolomn,showpacs,prl,amsmath,amssymb]{revtex4}
\usepackage{graphicx}
\usepackage{dcolumn}
\usepackage{bm}
\usepackage{amsmath}

\begin{document}

\preprint{APS}

\title{Mandala Networks: ultra-robust, ultra-small-world and highly sparse graphs}

\author{Cesar I. N. Sampaio Filho$^1$, Andr\'{e} A. Moreira$^1$, Roberto F. S. Andrade$^2$, Hans
  J. Herrmann$^{1,3}$, Jos\'{e} S. Andrade Jr.$^{1,3}\footnote{Correspondence to: soares@fisica.ufc.br}$}

\affiliation{$^1$Departamento de F\'{i}sica, Universidade Federal do Cear\'a, 
  60451-970 Fortaleza, Cear\'a, Brazil\\
  $^2$ Instituto de F\'{i}sica, Universidade Federal da Bahia, 
  40210-340 Salvador, Bahia, Brazil\\
  $^3$ Computational Physics for Engineering Materials, IfB, ETH Zurich, 
  Schafmattstrasse 6, 8093 Zurich, Switzerland}

\date{\today}

\begin{abstract}
  The increasing demands in security and reliability of
  infrastructures call for the optimal design of their embedded
  complex networks topologies. The following question then arises:
  what is the optimal layout to fulfill best all the demands? Here we
  present a general solution for this problem with scale-free
  networks, like the Internet and airline networks. Precisely, we
  disclose a way to systematically construct networks which are
  100$\%$ robust against random failures as well as to malicious
  attacks. Furthermore, as the sizes of these networks increase, their
  shortest paths become asymptotically invariant and densities of
  links go to zero, making them ultra-small worlds and highly sparse,
  respectively. The first property is ideal for communication and
  navigation purposes, while the second is interesting economically.
\end{abstract}

                              
\maketitle

The tremendous increase in complexity of infrastructural networks,
like the Internet and those related with transportation and energy
supply, is mandatorily accompanied by requirements of higher standards
of system reliability, security and robustness. This trend can only be
sustained if these complex networks have the right structure. Under
this framework, the scale-free property present in many real networks
determines important aspects related with their functionality
\cite{barabasiRevModPhys2002,newmanSiam2003,lopezPRL2005,boccaletti2006,barabasiScience2009}.
However, while scale-free networks are usually quite robust against
random failures, they typically break down rapidly under malicious
attacks
\cite{barabasiNature2000,callawayPRL2000,cohenPRL2000,cohenPRL2001,
  holmePRE2002, tanizawaPRE2005}. Numerical studies have recently
revealed that this weakness can be mitigated if their structure
becomes onion-like, which means that nodes of equal degree are
connected among each other and to nodes of higher degree
\cite{schneiderPNAS2011,schneiderJStat2011}. Since then, the
properties of onion-like structures have been extensively investigated
\cite{holmePRE2011,buonoEPL2011,schneiderEPL2012,stanleyPRE2012,zengPRE2012,dirkNature2013,dongPRE2013,havlinPRE2013,wangEPL2013,louzada2013}.
Based on this insight, here we will address the challenge of providing
a paradigm for complex networks with better topology. More precisely,
we show that it is possible to design a family of scale-free networks
which are completely robust to random failures as well as to malicious
attacks. Additionally, these networks also exhibit other improved
properties, like a finite shortest path and extreme sparseness in the
thermodynamic limit, which substantially increases communication and
reduces costs. Thus these new networks become potential candidates for
the design and implementation of complex infrastructural networks.

\begin{figure}[htb]
\includegraphics*[width=\columnwidth]{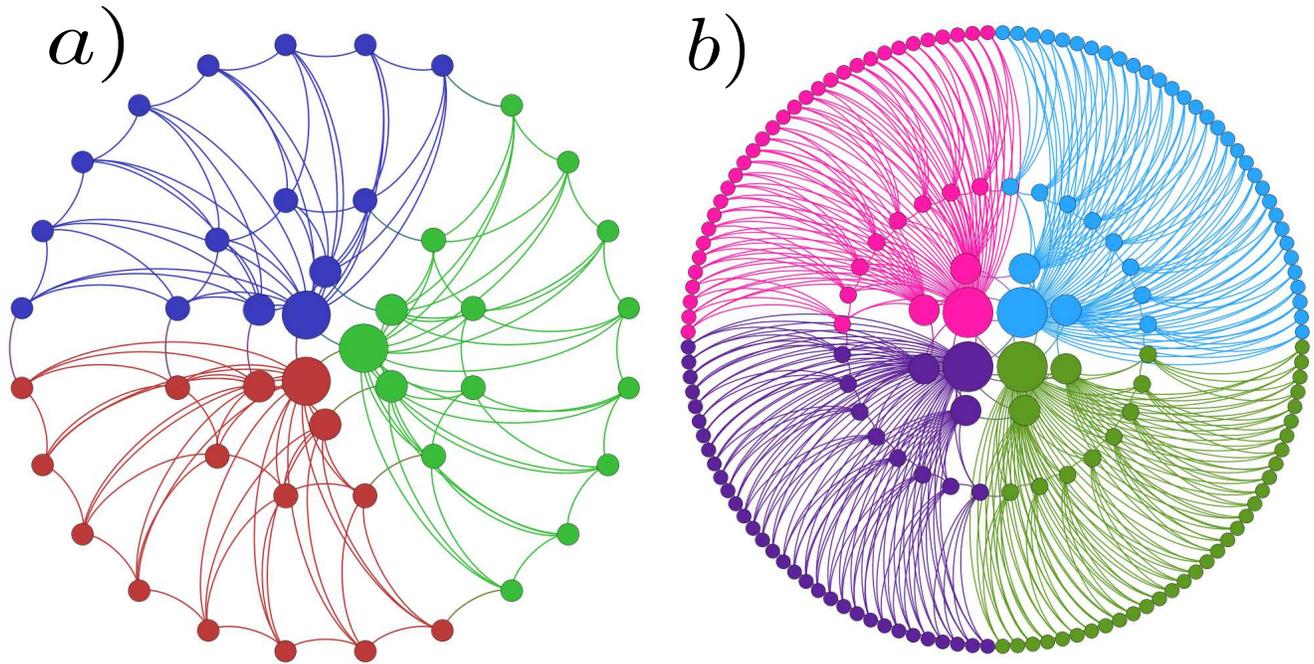}
\caption{(Color online) (a) Representation of the mandala network of
  type $A$, generated with parameters $b=2$, $n_{1}=3$ and
  $\lambda=2$.  The nodes correspond to circles whose areas are
  proportional to degree, and nodes in the same community have the
  same color. The first generation consists of a complete graph with
  three nodes defining the nucleus of the network.  To each one of
  these nodes, two new nodes are connected to form a connected
  circular ring of six nodes, corresponding to the second shell. Next,
  the most external nodes generate two new nodes forming a circular
  ring with twelve nodes (third shell). Every node in the same
  community is connected with all its ancestral ones. (b) Network of
  type $B$, generated with parameters $b=4$, $n_{1}=4$ and
  $\lambda=2$.}
\label{fig01}
\end{figure}
In the deterministic network model introduced here, the nodes
belonging to a given shell have intra-shell and inter-shell
connections, and the most connected nodes (hubs) are localized in the
innermost shells. The network is recurrently expanded in such a way
that every new generation corresponds to the addition of a new shell.
\begin{figure}[htb]
\includegraphics*[width=\columnwidth]{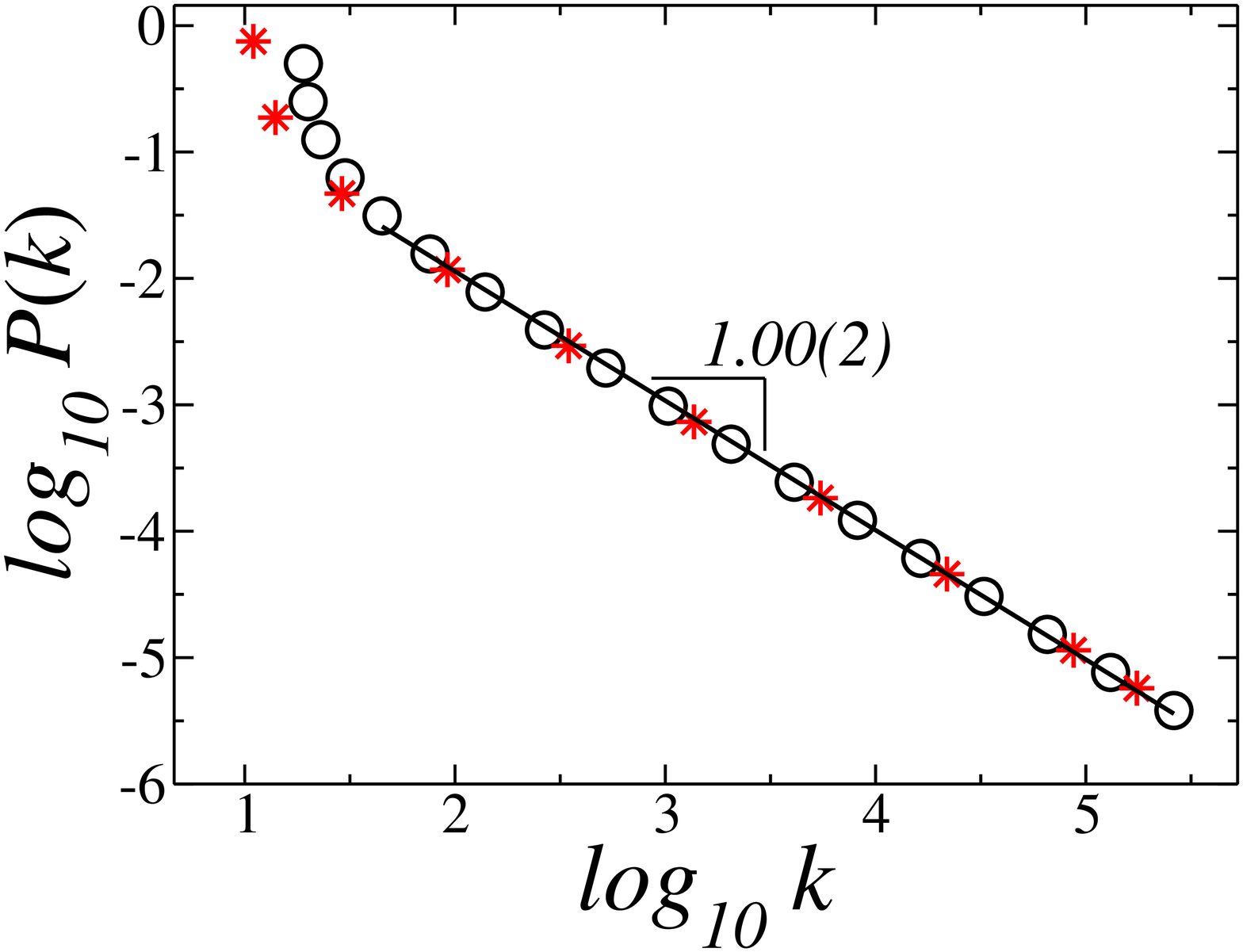}
\caption{(Color online) Logarithmic plot of the cumulative degree
  distribution for the networks $A$ (black circles) and $B$ (red
  stars). The solid line represents the least-squares fit to data in
  the scaling regions of a power law, $P(k)\sim k^{-\beta}$, with
  $\beta=1.00\pm 0.02$, which confirms our analytical result.}
\label{fig02}
\end{figure}
Examples of these networks with four shells are shown in
Figs.~\ref{fig01}(a) and (b). Here we coin the name {\it mandala
  network} for this new family of graphs. In the first case,
thereafter called network $A$, the first generation consists of a
nucleus with three central nodes forming a complete graph (first
shell). From each node in this nucleus, two new nodes emerge to form a
connected ring of six nodes, composing the second shell of the second
generation network. Following this iterative process, the third shell
in the third generation network has an additional connected ring with
twelve nodes, which, at this point, must also be linked to their
respective ancestral nodes in the first and second shells. The same
rules then apply for all new shells present in higher generation
networks. This design therefore imposes that nodes at the same shell
have the same degree. More precisely, the degree $k_{ig}$ of a node at
the $i$-$th$ shell in the $g$-$th$ network generation is given by,
\begin{equation}
k_{ig} = 2^{g-i+1} + (i-1).
\label{degreeEquation}
\end{equation}
Defining $n_{i}$ as the number of nodes in the $i$-$th$ shell, by
construction, we have that $n_{i+1}=2n_{i}$. From this relation, the
number of nodes in the network is given by $N=\sum_{i=1}^{g}n_{i}$,
where the summation is over the total number $g$ of shells.

In fact, the network described so far in Fig.~\ref{fig01}(a) is a
particular case resulting from the recursive method proposed here to
generate an ensemble of ultra-robust networks. For example, in
Fig.~\ref{fig01}(b) we show another example of mandala network,
thereafter named network $B$. Precisely, the method depends on a set
of three parameters, $(n_{1},b,\lambda)$, namely, the number of nodes
in the first generation, $n_{1}$, the number of new nodes added to
each node in the more external shell, $b$, and the scale factor,
$\lambda$, for node degree in successive generations. For instance,
the networks $A$ and $B$ are completely defined by the sets
$(3,2,2)$ and $(4,4,2)$, respectively.

\textbf{Results}

\textbf{Scale-free networks.} Any network generated by this
method has discrete degree spectra. In order to characterize the scale-free
dependence, we consider the cumulative degree distribution, $P(k)=\sum_{k^{'}\ge k}n(k^{'})/N$.
Taking into account that in each shell all nodes have the same degree,
the cumulative distribution can be written as,
$P(k_{ig})=\sum_{j=1}^{i}n_{j}/N$. Applying Eq.~(\ref{degreeEquation})
and the relation $n_{j+1}=bn_{j}$, it can be shown that the cumulative
distribution decays in the form, $P(k_{ig})\sim 1/k_{ig}$. In
Fig.~\ref{fig02}, we show a logarithmic plot of the cumulative degree
distribution for networks $A$ and $B$. In both cases, we have the same
scale-free dependence. At this point, an explanation about the
exponent of the degree distribution becomes necessary
\cite{andradePRL2009,Guo2010,munganPRL2011}.  By definition of our
network, the discrete degree distribution, $p(k)$, can be described
by, $p(k_{ig})\equiv n_{i}/N$, for a large number of shells.
Considering that $n_{i}\sim 2^{i}$, we have $p(k_{ig})\sim 1/k_{ig}$.
Therefore, the cumulative and probability distributions scale in the
same form. However, if we choose to work with binned intervals between
consecutive degrees, the degree distribution is calculated as,
$\overline{p}(k_{ig})\equiv n_{i}/N\Delta k_{ig}$, where $\Delta
k_{ig}=k_{ig}-k_{(i+1)g}$ is the width of the interval.  As $\Delta
k_{ig}\sim k_{ig}$, it is possible to show that then
$\overline{p}(k_{ig})\sim 1/k_{ig}^{2}$.

\begin{figure}[htb]
  \includegraphics*[width=\columnwidth]{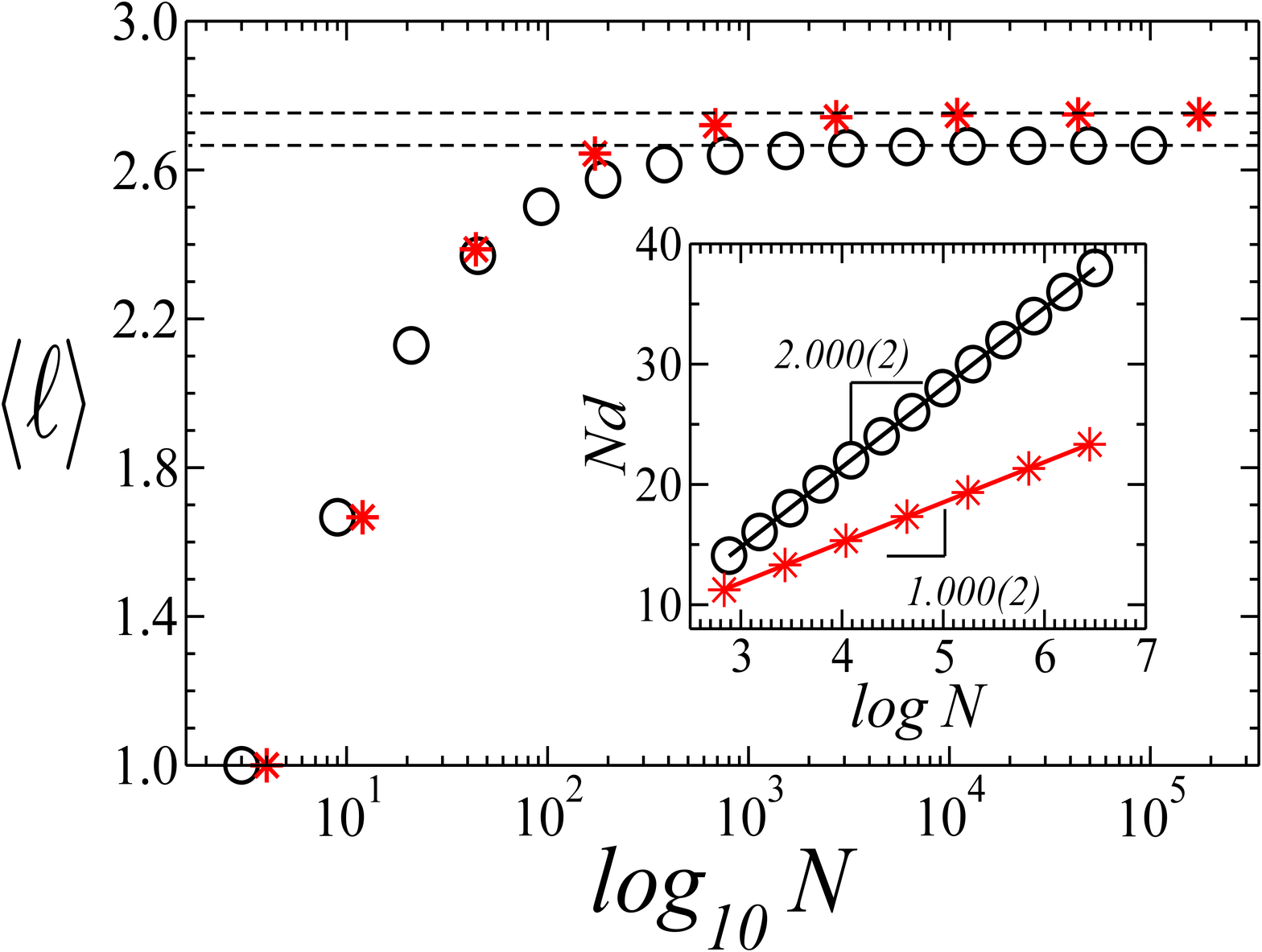}
  \caption{(Color online) Semi-log plot showing the dependence of the
    mean shortest path length $\left<\ell\right>$ on the number of
    nodes $N$, for the networks $A$ (black circles) and $B$ (red
    stars). As depicted, the mean shortest-path lengths of $A$ and $B$
    converge to the values $8/3$ (top dashed line) and $11/4$ (bottom
    dashed line), respectively, in the limit of a large number of
    nodes. Therefore, both networks can be considered as ultra-small
    worlds. The inset shows the semi-log plot of the density of
    connections $d$ as a function of the number of nodes $N$ in
    log-linear scale. Our analytical results reveal that
    $d\sim\frac{1}{N}\log N^{2}$ (black circles) for network $A$,
    while network $B$ behaves as $d\sim\frac{1}{N}\log N$ (red stars).
    The solid lines are the best fits to the numerically generated
    data sets, confirming these predicted behaviors. Hence both
    networks are highly sparse.}
\label{fig03}
\end{figure}

\textbf{Ultra-small-world networks.} Another important property of the mandala networks relates to the mean
shortest path length
$\left<\ell\right>=\sum_{ij}^{N}\ell_{ij}/\left[N(N-1)\right]$, where
$\ell_{ij}$ is the shortest distance between any two nodes $i$ and $j$
in the network, and the summation goes over all possible node pairs in
the system. In our case, this expression can be written in a more
convenient form as,
\begin{equation}
\left<\ell \right> =\frac{1}{N(N-1)}\sum_{j=1}^{g}n_{j}\phi_{j},
\label{shortestPathEq1}
\end{equation}
where $\phi_{j}=\sum_{k=1}^{N}\ell_{jk}$ is the sum of the shortest
path lengths connecting a node in the $j$-$th$ shell with all other
nodes in the network, $n_{j}$ is the number of nodes in the $j$-$th$
shell, and the summation goes over the number of shells. Using the
symmetry of the network $A$, for example, it is possible to show that
$\phi_{j}=\alpha_{j}N+\xi_{j}$, where $\xi_{j}$ has different values
for different shells, and $\alpha_{i}$ is given by $5/3$, $7/3$,
$5/2$, $31/12$, $63/24$, for $i=1, 2, 3, 4$, and $5$, respectively, so
that $\alpha\to 8/3$, for $i \to \infty$. Taking into account the
linear dependence of $\phi_{j}$ with $N$ and considering the relations
for $n_{j}$, Eq.~(\ref{shortestPathEq1}) reduces to
\begin{equation}
\left<\ell \right>=\alpha+\frac{O(N)}{N^{2}},
\label{shortestPathEq02}
\end{equation} 
which leads to $\left<\ell \right>\to 8/3$ in the thermodynamic limit,
$N\to\infty$. We show in Fig.~\ref{fig03} a semi-log plot of the mean
shortest path length as a function of the number of nodes. The
asymptotic convergence confirms our analytical result and therefore
indicates that our network has an ultra-small-world behavior, namely
$\left<\ell\right>$ becomes independent of $N$. One should note,
however, that this result is still different from the case of a
complete graph, for which $\left<\ell\right>=1$, corresponding to the
mean-field limit.  Applying a similar sequence of calculations to the
network $B$, it can be readily shown that the mean shortest path
length for this topology also converges to a constant in the limit of
large system sizes, but now equal to $11/4$.

\textbf{Highly sparse graphs.} Next, we define the density $d$ of connections as the ratio between
the number of existing connections and the maximal number of possible
connections for an undirected network with $N$ nodes,
$d=\sum_{i}n_{i}k_{ig}/\left[N(N-1)\right]$. Considering the
expression for $k_{ig}$ given by Eq.~(\ref{degreeEquation}), we can
rewrite the definition of $d$ in the following way:
\begin{equation}
  d = \frac{1}{N(N-1)}\left[ \sum_{i}^{g}n_{i}2^{g-i+1}+\sum_{i}^{g}n_{i}(i-1)\right].
  \label{eq03Density}
\end{equation}
Expressing both summations in Eq.~(\ref{eq03Density}) in terms of the
number of nodes in the network, $N=\sum_{i}^{g}n_{i}$, and considering
the limit of a very large number of generations, we obtain,
\begin{equation}
  d \sim \frac{1}{N}\log N^{2}.
\label{eq04Density}
\end{equation} 
The inset of Fig.~\ref{fig03} shows the dependence of the density of
connections on the number of nodes for networks of type $A$,
confirming the asymptotic behavior predicted by
Eq.~(\ref{eq04Density}). In the case of network $B$, where the number
of new nodes generated is twice that of network $A$, the density of
connections decays faster. Indeed, applying the same approach and
considering $b=4$, it is possible to show for network $B$ that $d \sim
\left(\log N\right)/N$. As a consequence, we conclude that our
networks, despite of their ultra-small-world property, are extremely
sparse when compared to the behavior of a complete graph, $d=1$, and
has only logarithmic correction to $d\sim 1/N$ that is valid for the
Erd\"os-R\'enyi network at the percolation threshold.
\begin{figure}[htb]
\includegraphics*[width=\columnwidth]{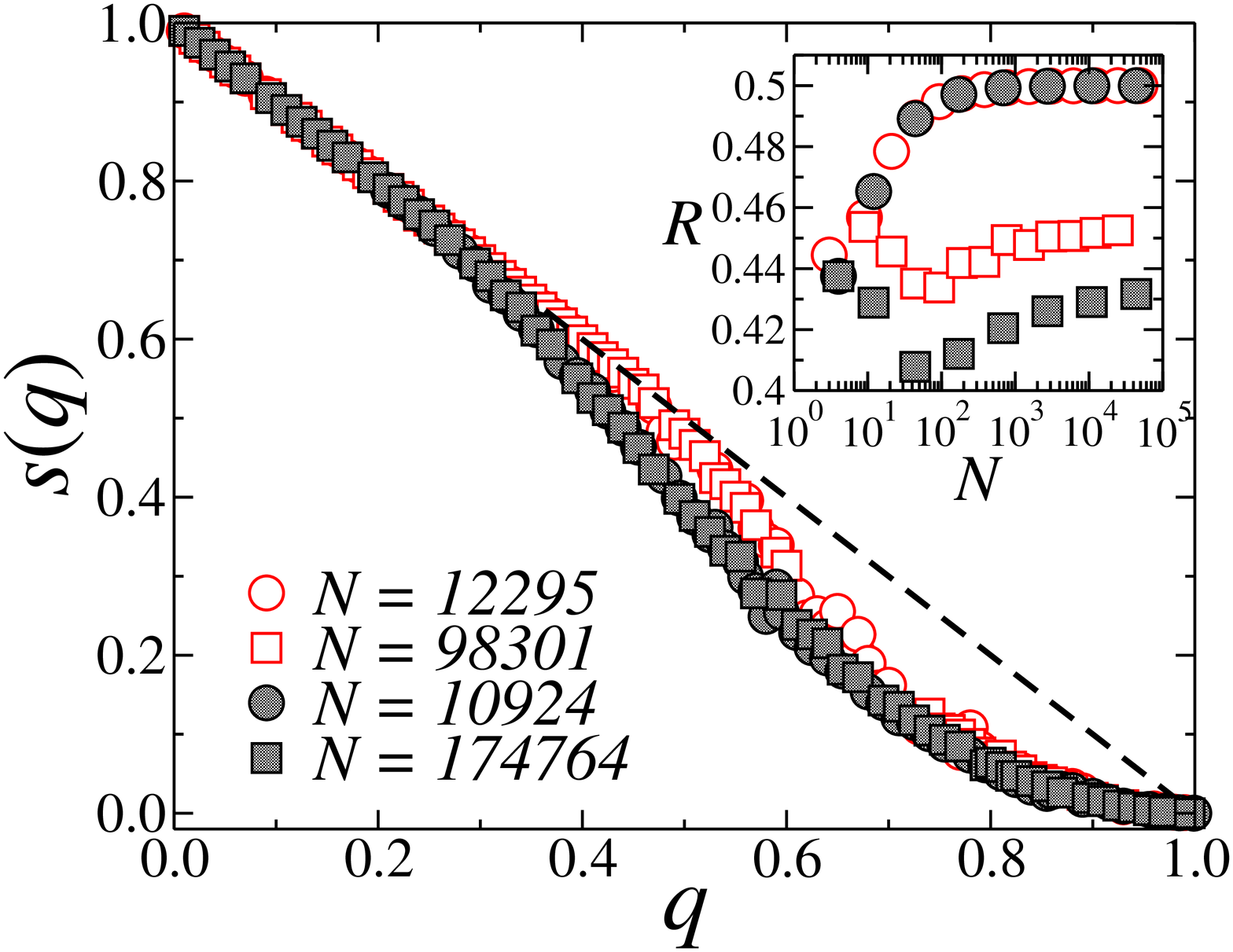}
\caption{(Color online) Fraction $s(q)$ of nodes belonging to the
  spanning cluster as a function of the fraction of removed nodes $q$,
  for malicious attacks targeted by degree (dashed line) and random
  removals (symbols) of nodes. In the case of malicious attacks, an
  identical behavior is observed for sufficiently large networks,
  $s=1-q$, regardless of the type ($A$ or $B$) of the network
  used in the calculations. In the random-attack case, our results
  exhibit high robustness for both $A$ and $B$ networks, that are also
  shown to be practically independent of the network size for the
  different numbers of nodes $N$ considered. The inset shows the
  robustness measure $R$ as a function of $N$ for networks $A$ and $B$
  subjected to malicious attacks targeted by degree (open symbols) as
  well as random failures (closed symbols). An asymptotic convergence
  towards the maximum robustness value, $R_{max}=1/2$, is observed for
  the case of malicious attacks, regardless of the network type. Both
  networks are less resilient to random attacks than to targeted ones,
  but still rather robust as compared to other models and real
  networks \cite{schneiderPNAS2011}, with $R\approx 0.45$ and $0.43$,
  for types $A$ and $B$, respectively.}
\label{fig04}
\end{figure}

\textbf{Ultra-robust graphs.} The framework of percolation is usually considered for the analysis of
the robustness of complex networks
\cite{callawayPRL2000,newmanPRE2001,cohenPRL2003,moreiraPRL2009,moreiraPRE2010,newmanNature2011,peixotoPRL2012,taylorEPL2012,wangEPL2013}.
In this context, robustness is typically quantified by the critical
fraction $q_c$ of removed nodes that leads to a total collapse of the
network
\cite{barabasiNature2000,cohenPRL2001,holmePRE2002,schneiderPNAS2011}.
Nevertheless, as previously reported
\cite{schneiderPNAS2011,schneiderJStat2011,holmePRE2011,schneiderEPL2012},
this approach does not account for situations in which the system can
suffer a big damage without breaking down completely. The size of the
giant component during the removal process of nodes has been recently
introduced \cite{schneiderPNAS2011} as a new measure to robustness,
\begin{equation}
R=\frac{1}{N+1}\sum_{Q=1}^{N}s,
\label{eq03}
\end{equation}
where $s$ is the fraction of nodes belonging to the giant component
after removing $Q=qN$ nodes, $q$ is the fraction of nodes removed, and
$R$ is in the range $\left[0,1/2\right]$. The limit $R=0$ corresponds
to a system of isolated nodes, while $R=1/2$ to the most robust
network, which is the case, for example, of a completely connected
graph. Here we check the robustness of our complex network model when
subjected to both mechanisms of malicious and random attacks
\cite{barabasiNature2000,callawayPRL2000,cohenPRL2000,cohenPRL2001,holmePRE2002}.

Considering that the targets of malicious attacks are the surviving
nodes with highest degree, in our network model, the process starts by
selecting one of the nodes of the first shell. The deletion of such
node will not cause the removal of other nodes, since any node in any
shell has links to nodes in the previous and subsequent shells.  Thus,
the number of nodes of the giant component decreases only by one. The
same occurs until all nodes in the first shell are erased.  The
attacks are then directed to the nodes in the second shell, but again
all nodes in the second shell are connected to nodes in the next
shells, so that the same argument applies. Thus, it follows that,
$s=1-q$, which is valid up to the point when the remaining nodes are
located in the last shell. This behavior is confirmed by results of
numerical simulations shown in the main plot of Fig.~\ref{fig04}
(dashed line), regardless of the size and the type $A$ or $B$ of the
network. It shows further that the integrity of the giant component is
maintained up to nearly its total breakdown for either malicious or
random strategies of node removal. Interestingly, the results in
Fig.~\ref{fig04} also indicate that random attacks are generally more
efficient than malicious ones.

From the behavior of $s$ and Eq.~(\ref{eq03}), the robustness $R$ to
malicious attacks can be calculated as,
$R=\sum_{i}^{N-1}(N-i)/(N+1)=(2N+1)/2N$. As shown in the inset of
Fig.~\ref{fig04}, $R$ converges asymptotically towards the maximum
value, $R_{max}=1/2$, as expected for the case of malicious attacks to
sufficiently large networks $A$ and $B$. Mandala networks of different
types and sizes also behave similarly when subjected to random
attacks. Nevertheless, even for random failures, the level of
robustness for our networks is superior when compared to other
networks \cite{schneiderPNAS2011}, with $R\approx 0.45$ for type $A$,
and $R\approx 0.43$, for type $B$.
\begin{figure}[htb]
\includegraphics*[width=\columnwidth]{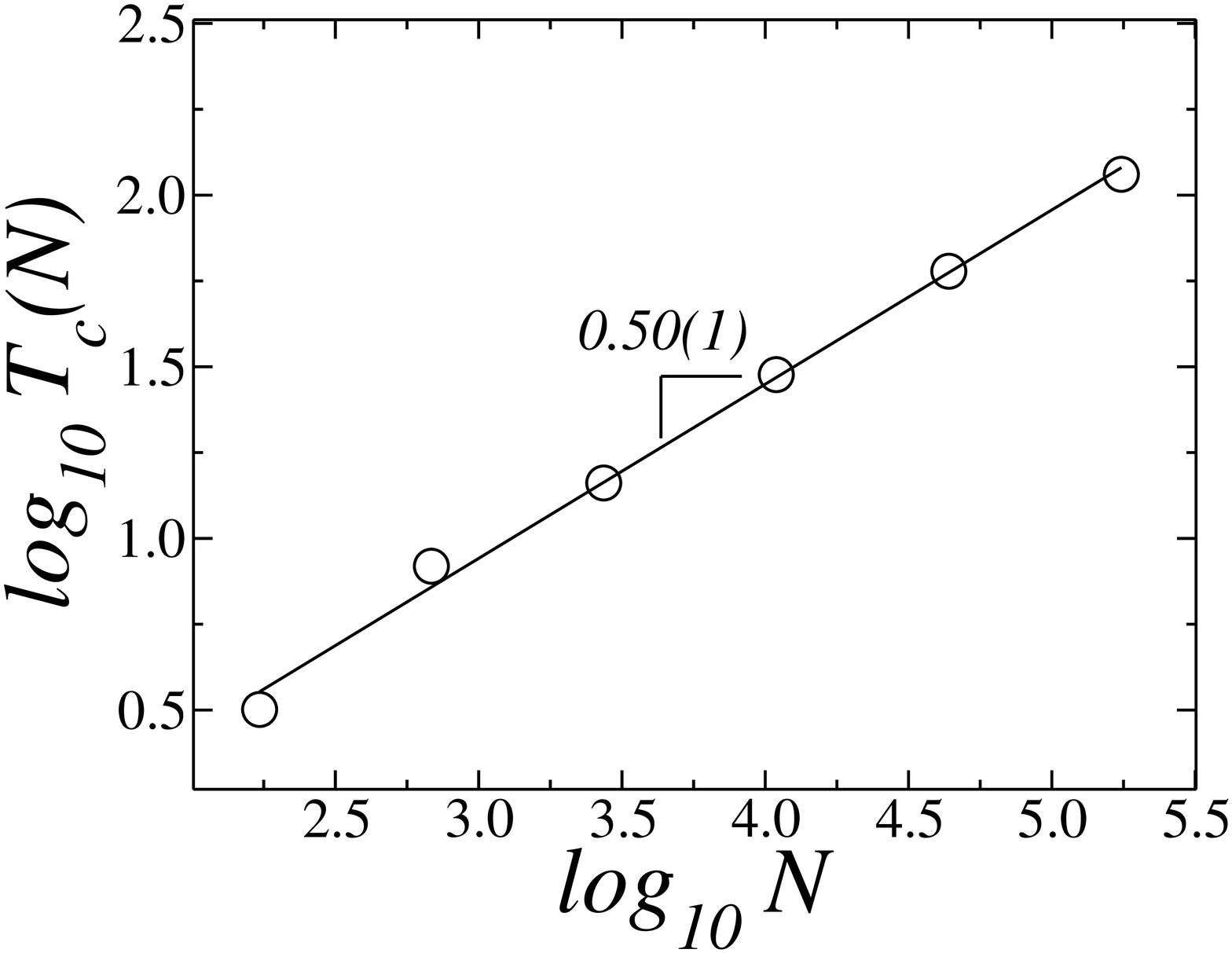}
\caption{Log-log plot showing the finite-size scaling analysis for the
  critical temperature of the Ising model implemented on the network
  model $A$. The critical temperature diverges according to a
  power-law dependence, $T_{c}(N)\sim N^{1/\overline{\nu}}$, with
  $1/\overline{\nu} = 0.50\pm 0.01$.}
\label{fig05}
\end{figure}

\textbf{The Ising model.} In small-world networks, the fact that the diameter of the graph does
not grow faster than $\log N$ implies an infinite dimension.
Mean-field theories therefore can successfully describe their critical
behavior
\cite{newmanPRE1999,barabasiRevModPhys2002,indekeuPRL2005,mendesRevModPhys2008,satorrasPRL2008,satorrasPRE2011}.
In order to investigate how collective ordering emerges in mandala
networks, for which the shortest-path length is independent on $N$, we
consider Ising spins $\sigma_{i}$ associated to their nodes and
ferromagnetic interactions $J$ between them on the edges. Adopting the
reduced Hamiltonian, ${\cal
  H}/k_{b}T=-J\sum_{ij}\sigma_{i}\sigma_{j}$, we perform Monte Carlo
(MC) simulations on networks of type $A$ for different system sizes
$N$ and temperature $T$ values. In particular, we analyse the
finite-size scaling properties of the model at the critical
temperature. The results in Fig.~\ref{fig05} show that the divergence
of the finite-size critical temperature with $N$, measured from the
peak of the susceptibility, has the form, $T_{c}(N)\sim
N^{1/\overline{\nu}}$, with a critical exponent,
$1/\overline{\nu}=0.50\pm 0.01$.

\textbf{Discussion}

In summary, we have presented a recursive method to generate an
ensemble of ultra-robust networks defined by a set of three
parameters, namely $(n_{1},b,\lambda)$. We have shown that the
networks originated from our model have scale-free topologies and are
ultra-robust either against malicious attacks or random failures of
nodes. From analytical calculations confirmed through numerical
simulation results, we have demonstrated that these networks are also
ultra-small, i.e., the average shortest-path lengths of sufficiently
large networks become independent of their number of nodes.
Surprisingly, our results also show that, as compared to a complete
graph, which is also ultra-small and ultra-robust, these networks are
highly sparse, with the density of edges going to zero with system
size. Finally, we have verified that the critical temperature of the
Ising model on the mandala network topology diverges with system size
according to a power-law dependence, described by an exponent
$1/\overline{\nu}=0.50\pm0.01$. We expect to generalize this last
result to other universality classes, for example, considering
directed percolation and self-organized models on our deterministic
networks.

\textbf{Methods}

\textbf{The Ising model.} The Monte Carlo simulations of the
Ising model on the mandala networks were performed using the Metropolis
algorithm, starting from different initial spin configurations. In order to
study the critical behavior of the system, we considered the magnetization 
$M_L$ and the susceptibility $\chi_L$, which are defined by

\begin{equation}
 M_{N}(T) = \left<\left< m \right>_{time} \right>_{sample},
\label{eq1Methods}
\end{equation}

\begin{equation}
 \chi_{N}(T) = N \left[\left< \left< m^{2} \right>_{time}  - \left< m \right>_{time}^{2} \right>_{sample}\right],
\label{eq2Methods}
\end{equation}

where $\left<m\right> = |\frac{1}{N}\sum_{i=1}^{N}\sigma_{i}|$, $T$ is the temperature
and $N$ is the total number of nodes in network. The symbols $<\cdots>_{time}$ and $<\cdots>_{sample}$, respectively,
denote time averages taken in the stationary state and configurational averages taken over $100$ independent
samples. Time is measured in Monte Carlo steps (MCS), and $1$ MCS corresponds to $N$ attempts of changing the states
of the spins. In our simulations, the initial $10^5$ MCS were utilized to guarantee the system reached the steady state,
after which the time averages were estimated using the next $6\times 10^5$ MCS.The value of temperature where $\chi_{N}$
has a maximum is identified as $T_{c}(N)$ for $N = 172, 684, 2732, 10924, 43692,$ and $174764$.

\textbf{Acknowledgments}

We thank the Brazilian agencies CNPq, CAPES, FUNCAP, and FAPESB, the
National Institute of Science and Technology for Complex Systems
(INCT-SC Brazil), to the ETH Risk Center (Switzerland), and the
European Research Council through Grant FlowCSS No.~FP7-319968 for
financial support.

\textbf{Authors contributions}

The authors C.S.F., A.M., R.A., H.J.H., and J.S.J., contributed equally. 

\textbf{Additional information}

Competing financial interests: The authors declare no competing financial interests.

\end{document}